\documentclass[10pt]{article}

\usepackage{amsmath}
\usepackage{amssymb}  
\usepackage[T1]{fontenc}
\usepackage{pifont}
\usepackage{pstricks}
\usepackage{amsfonts}
\usepackage{graphicx}
\usepackage{amsmath}
\usepackage{authblk}
\usepackage{pstricks} 
\usepackage{pstricks-add}
\usepackage{caption}
\usepackage{placeins}
\usepackage{pstricks}
\usepackage{pdfpages}
\usepackage{framed}
\usepackage{cancel}
\usepackage{bm}
\usepackage{epstopdf}
\usepackage{tikz}
\def\rr2dot{\mathop{\bf r}\limits}
\def\x2dot{\mathop{x}\limits}
\def\y2dot{\mathop{y}\limits}
\def\h2dot{\mathop{h}\limits}

\def\bfy2dot{\mathop{\bf y}\limits}
\def\z2dot{\mathop{z}\limits}

\def\csi2dot{\mathop{\xi}\limits}
\def\et2dot{\mathop{\eta}\limits}
\def\bet2dot{\mathop{\beta}\limits}
\def\t2dot{\mathop{\theta}\limits}
\def\s2dot{\mathop{\sigma}\limits}
\def\d2dot{\mathop{\delta}\limits}
\def\q2dot{\mathop{q}\limits}
\def\l2dot{\mathop{\lambda}\limits}
\def\ps2dot{\mathop{{\cal E}}\limits}
\def\tet2dot{\mathop{\theta}\limits}
\def\bfx2dot{\mathop{\bf x}\limits}
\def\bfy2dot{\mathop{\bf y}\limits}
\def\bfq2dot{\mathop{\bf q}\limits}
\def\bfr2dot{\mathop{\bf r}\limits}
\def\bbfq2dot{\mathop{\bar {\bf q}}\limits}
\def\w2{\mathop{W}\limits}
\def\xgrande2dot{\mathop{\bf X}\limits}
\def\p02dot{\mathop{P}\limits}
\def\a2dot{\mathop{A}\limits}

\newtheorem{prop}{Proposition}
\newtheorem{propr}{Property}
\newtheorem{rem}{Remark}
\newtheorem{exe}{Example}

\title{Is the ${\check {\rm C}}$etaev condition inferable?}
\author{F.~Talamucci}
\affil{{\it DIMAI, Dipartimento di Matematica e Informatica ``Ulisse Dini''},\\
{\it	Universit\`a degli Studi di Firenze, Italy}\\
{\it	e-mail: federico.talamucci@unifi.it}}
\date{}

\begin{document}
	\bibliographystyle{plain}
	
	\setcounter{equation}{0}

	\maketitle
	
	\vspace{.5truecm}
	
	
	\vspace{.5truecm}
	
	
	\vspace{.5truecm}
	

\begin{abstract}
	
\noindent	
In the context of holonomic constrained systems the identification of virtual displacements is clear and consolidated: this gives the possibility, once the class of displacements have been combined with Newton's equations, to write the correct equations of motion for the constrained system. The method combines d'Alembert principle with the Lagrange formalism.
As far as nonholonomic constraints are concerned, the conjecture that dates back to ${\check {\rm C}}$eteav actually defines a class of virtual displacements through which the method d'Alebert--Lagrange can be applied again.
Much literature is dedicated to the ${\check {\rm C}}$etaev rule from both the theoretical and experimental points of view.
The absence of a rigorous (mathematical) validation of the rule inferable from the constraint equations has been declared expired in a recent publication: our main objective is a critical investigation of the stated result.
\end{abstract}

\section{Introduction}

\noindent
Our study concernes discrete systems constrained via nonholonomic constraints. 
Among the various approaches used to deal with the problem (virtual displacements, variational principles in difeerential or in integral form, algebraic and geometric methods) our attention is directed to the d'Alembert--Lagrange principle (d'A--L P.~). The classic formulation of the principle will be presented in the next Section, while here we address the issue from a substantial and qualitative point of view.

\noindent
As it is known (see for instance \cite{gant} or \cite{pars}), the formulation of d'A--L P.~takes place starting from Newton's equations of motion, selecting a particular category of displacements $\delta {\bf r}$ (we will specify the notations later on) which must correspond to the dynamics (possibly constrained) of the system.
The situation is clear and universally accepted when we are dealing with systems bound by geometric conditions (holonomic constraints), a case in which the displacements are distinctly identifiable and spontaneously linked to simple concepts of geometry (tangent space of a manifold).

\noindent
The approach through d'A--L P.~is certainly not absent in literature when we move on to the more complex category of kinematic constraints, i.~e.~constraints which involve the velocities of the system and which cannot be reduced  to geometric constraints by integration: starting from the fundamental text \cite{neimark}
the problem of how to define displacements is widely recurring in literature, although the alternative approaches we just mentioned above perhaps prevail in the study of nonholonomic systems. We refer to \cite{li} for 
an useful review and an updated state of the art on virtual displacements in nonlinear nonholonomic constraints.

\noindent
In comprehending the role and potentiality of the d'A--L P.~when dealing with nonholonomic systems, there is a very clear dividing line of demarcation dividing the linear case and the nonlinear one: 
if in the first case the application of the method  d'A--L P.~is a natural and simple extension of the holonomic model, in the latter case the task of assigning dispacements to the system is much more difficult and unclear.
From the mathematical point of view the question is very simple to introduce: the matter is to deduce appropriate integer conditions in terms of $\delta {\bf r}$ starting from a set of constraints on the state $({\bf r}, {\dot {\bf r}})$ of the system, which cannot be integrated.
Nevertheless, it must be said that the topic of nonlinear kinematic constraints is quite problematic and is way beyond the perspective of the d'A--L P.~we are focussing on: actually, starting from the very existence of physical models of this type, the debating issues concern the concrete feasibility of physical models  \cite{benentipendulum}, the distinct role from control forces \cite{zek1}, the axiomatic and theoretical features that provide a valid generalization of the linear case \cite{flan0}, the mathematical aspect of expressing a certain condition in multiple equivalent ways and in different sets of coordinates \cite{zek2}, just to mention a few questions.
In addition to the references just above, we indicate the texts \cite{lurie}, \cite{papastrav} and the articles \cite{benenti1}, \cite{borisov}, \cite{leon1}, \cite{marle}, \cite{mei} which examine in depth the topic of nonlinear constraints from the historical and axiomatic point of view.

\noindent
Coming back to the themes of our work, let us say that our analysis is mainly aimed at examining a condition that establishes the class of displacements $\delta {\bf r}$ in a very simple way and which is known as ${\check {\rm C}}$etaev rule (\cite{cetaev}, we also refer to \cite{rumycet}, \cite{rumy}).
Naturally, as we will recall later, the rule overlaps exactly with the existing conditions used for the holonomic case and the linear kinematic case. The interesting aspect is that historically the hypothesis was born with the aim of making the d'A--L P.~and the Gauss Principle equivalent, in order to achieve agreement through the Hertz--Helder principle.

\noindent
Going through the specific literature, we understand that the ${\check {\rm C}}$etaev rule is in the state of a postulate, that is, there does not seem to be a rigorous theoretical justification starting from the laws of the mechanics \cite{li}.
At the same time, the debate is also open in the context of experimental tests
and there are conflicting opinions even when the hypothesis is tested directly with experimental procedures \cite{lewis}, \cite{xuli}.
For this reason we were impressed by the content of \cite{flan}, where a full theoretical argument intends to extend the validity of the d'A--L P.~to treat general nonholonmic systems: essentially, the ${\check {\rm C}}$etaev rule is claimed as deriving (through mathematical steps) directly from the constraint equations and in this way the displacements defined by the rule itself are fully justified.
The procedure is also extended to higher order constraints, which involve the accelerations of the system.

\noindent
The dissertation in \cite{flan} is undoubtedly outstanding and exhaustive, since it explores the most remarkable aspects of the theory on nonholonomic constraints (including various types of Principles, the debated transpositional rule, the pseudovelocity formalism). This is why we find it appropriate to dwell on the  mathematical tools by which the rule seems to find its demonstration.

\noindent
At this point, it seems appropriate to us for the sake of clarity to present the structure of the rest of the present work and indicate the main steps:
\begin{itemize}
	\item[$(i)$] The first part of Section $2$ contains ordinary but necessary notions to present the equations of the mechanical model; in the second part the ${\check {\rm C}}$etaev condition (eq.~(\ref{cetaev})) is introduced.
	In the last Paragraph of the same Section we briefly present the equations of motion for the constrained system written accordingly in assuming the ${\check {\rm C}}$etaev condition.

	\item[$(ii)$] In Section $3$ we focus on the condition and we critically explore the arguments and deductions presented in \cite{flan}, which should validate the hypothesis from a mathematical point of view.
	In the second Paragraph of the same Section we point out some aspects and consequences of the condition that have a precise physical meaning and that predispose the condition as one that naturally extends the theory of holonomic systems.
	
	\item[$(iii)$] Finally, in Section $4$ conclusions are summarized and future developments on the topic are declared.
\end{itemize}

\section{The mathematical model}

\subsection{The d'Alembert--Lagrange Principle}

\noindent
The general context on which we will make our considerations consists of a system of $N$ material points governed by the Newton's equations
\begin{equation}
	\label{newton}
	m_s {\bfr2dot^{..}}_s = {\bf F}_s+{\bm R}_s, \qquad s=1, \dots, N
\end{equation}
where ${\bf r}_s$ is the vector in ${\Bbb R}^3$ locating the $s$--th point, $m_s$ the mass, ${\bf F}_s$ is the corresponding active force and ${\bf R}_s$ is the unknown force (exerted on the $s$--th point) due to restrictions imposed to the system.
A variation of the configuration of the system at a fixed time which is consistent with the constraint is a 
virtual displacement $\delta {\bf r}_s$.
The standard formulation of the equations of motion using the d'A--L P.~passes through the definition of ideal constraint (we use the summation symbol $\Sigma$ for greater clarity, given that the range of indices is changeable)
\begin{equation}
\label{ic}
\sum\limits_{s=1}^N {\bm R}_s \cdot \delta {\bf r}_s=0
\end{equation} 
so that it can be written
\begin{equation}
	\label{dap}
\sum\limits_{s=1}^N \left(	m_s {\bfr2dot^{..}}_s - {\bf F}_s\right)\cdot \delta {\bf r}_s=0
\end{equation}
from which the correct equations of motion should be deduced.
If generalized independent coordinates $q_1$, $\dots$, $q_n$, $n=3N$ are used, then $\delta {\bf r}_s=
\sum\limits_{j=1}^{n}\dfrac{\partial {\bf r}_s}{\partial q_j}\delta q_j$ for each $s=1, \dots, N$ and (\ref{dap}) is  
\begin{equation}
	\label{dapl}
\sum\limits_{j=1}^n \left(
\dfrac{d}{dt}\dfrac{\partial T}{\partial {\dot q}_j}-\dfrac{\partial T}{\partial q_j}
-F^{(j)}\right) \delta q_j=0 
\end{equation}
where the kinetic energy $T=\dfrac{1}{2}\sum\limits_{s=1}^N m_s {\dot {\bf r}_s^2}$ is a function of the generalized coordinates ${\bf q}=(q_1, \dots, q_n)$ and the generalized velocities ${\dot {\bf q}}=({\dot q}_1, \dots, q_n)$ by means of  
\begin{equation}
	\label{rq}
{\bf r}_s={\bf r}_s  ({\bf q,t}), \qquad {\dot {\bf r}_s}({\bf q}, {\dot {\bf q}},t)=
\sum\limits_{j=1}^n \dfrac{\partial {\bf r}_s}{\partial q_j}{\dot q}_j+\dfrac{\partial {\bf r}_s}{\partial q_j}, \qquad s=1, \dots, N
\end{equation}
and $F^{(j)}=\sum\limits_{s=1}^N {\bf F}_s \cdot \dfrac{\partial {\bf r}_s}{\partial q_j}$ is the generalized force along the $j$--th direction, $j=1, \dots, n$.
The generalized constraint forces are $R^{(j)}=\sum\limits_{s=1}^N {\bf R_s}\cdot \dfrac{\partial {\bf r}_s}{\partial q_j}$ and in the lagrangian coordinates 
the condition (\ref{ic}) of ideal constraint is
\begin{equation}
	\label{iclagr}
	\sum\limits_{j=1}^n R^{(j)} \delta q_j=0.
\end{equation}
Still within the standard premises of the problem, we recall that if part of the force comes from a generalized potential in a way such that
$$
F^{(j)}_P=\dfrac{\partial U}{\partial q_j}-\dfrac{d}{dt}\dfrac{\partial U}{\partial {\dot q_j}}, \qquad 
		j=1, \dots, n
$$
then the introduction of the Lagrangian function ${\cal L}({\bf q}, {\dot {\bf q}}, t)=T+U$ makes us write (\ref{dapl}) as 
\begin{equation}
	\label{dapl1}
	\sum\limits_{j=1}^n \left(
	\dfrac{d}{dt}\dfrac{\partial {\cal L}}{\partial {\dot q}_j}-\dfrac{\partial {\cal L}}{\partial q_j}-F^{(j)}_{NP}\right) \delta q_j=0 
	\end{equation}
where the term $F^{(j)}_{NP}$ takes into account the remaining active forces not deriving from a potential.

\begin{rem}
It should be noted that in the passage from the cartesian coordinates ${\bf r}_s$ to the generalized ones 
no further geometric constraint occurs, i.~e.~$N=3n$: if this happens, the selection of a smaller number of generalized coordinates $n<3N$ does not involve substantial changes other than the reformulation of the constraint forces ${\bf R}_s$.
\end{rem}

\subsection{The constraint conditions}

\noindent
We assume that the restrictions can be formulated by the constraints equations
\begin{equation}
	\label{vinc}
	\psi_\nu ({\bf r}_1, \dots, {\bf r}_N, {\dot {\bf r}}_1, \dots, {\dot {\bf r}}_N, t)=0, \qquad \nu=1, \dots, k<3N.
\end{equation}
hence we consider kinematic constraints (linear or nonlinear) involving the positions and the velocities os the points.
The constraints are independent according to the following assumption:
\begin{eqnarray}
	\label{psiindep} \textrm{the}\;k\;\textrm{vectors} \;\dfrac{\partial \psi_\nu}{\partial {\dot {\bf r}}}\in {\Bbb R}^{3N},\; \nu=1, \dots, k &\textrm{are independent}
\end{eqnarray}
where ${\bm r}=({\bm r}_1, \dots, {\bm r}_N)\in {\Bbb R}^{3N}$; equivalently, the rank of the $(k\times 3N)$ jacobian matrix $\dfrac{\partial {\bm \psi}}{\partial {\dot {\bf r}}}$, ${\bm \psi}=(\psi_1, \dots, \psi_k)$, is full and has its maximum value $k$.

\noindent
By means of (\ref{rq}) the constraint equations can be written in terms of the generalized coordinates as
\begin{equation}
\label{gvinc}
g_\nu({\bf q}, {\dot {\bf q}},t)=0, \qquad \nu=1, \dots, k
\end{equation}
Owing to the chain relation $\dfrac{\partial {\bf g}}{\partial {\dot {\bf q}}}=
\dfrac{\partial {\bf \psi}}{\partial {\dot {\bf r}}}\dfrac{\partial {\dot {\bf r}}}{\partial {\dot {\bf q}}}=
\dfrac{\partial {\bf \psi}}{\partial {\dot {\bf r}}}\dfrac{\partial {\bf r}}{\partial {\bf q}}$, where ${\bf g}=(g_1, \dots, g_k)$, we see that the constraints (\ref{gvinc}) are independent too, by virtue of (\ref{psiindep}) and the non--singularity of $\partial {\bf r}/\partial {\bf q}$, since the generalized coordinates ${\bf q}$ are independent.

\noindent
The delicate and crucial point, still debated, can be summarized by the following question: what is the correct set of virtual displacements to consider so that (\ref{ic}) or equivalently (\ref{iclagr}) is valid?
Once this has been established, the correct equations of motion can be explicitly deduced from (\ref{dapl}) or (\ref{dapl1}).

\noindent
It is appropriate to consider the holonomic case as established and as an element of comparison for correctly generalizing the hypotheses: we recall that \cite{gant} in the case of geometric constraints of the type
\begin{equation}
	\label{vincol}
\phi_\nu ({\bf r}_1, \dots, {\bf r}_N, t)=0 \qquad \Longrightarrow \;\;f_\nu({\bf q},t)=0
\end{equation}
(see (\ref{rq})), the consolidated theory in this regard provides the following category of virtual displacements:
\begin{equation}
	\label{virtdisplol}
\sum\limits_{j=1}^n \dfrac{\partial f_\nu}{\partial q_j}\delta q_j=0
\end{equation}
It is worth noting that, setting the constraint in the form (\ref{gvinc}) by differentiation:
\begin{equation}
	\label{ec}
g_\nu({\bf q}, {\dot {\bf q}},t)={\dot f}_\nu({\bf q},t)=\sum\limits_{j=1}^n \dfrac{\partial f_\nu}{\partial q_j}{\dot q}_j+\dfrac{\partial f_\nu}{\partial t}=0
\end{equation}
then condition (\ref{virtdisplol}) coincides with
\begin{equation}
\label{cetaev}
\delta^{(c)} g_\nu=\sum\limits_{j=1}^n \dfrac{\partial g_\nu}{\partial {\dot q}_j}\delta q_j=0
\end{equation}
It is known that even extending from the holonomic case to that of linear kinematic constraints
\begin{equation}
	\label{vinclin}
	g_\nu ({\bf q}, {\dot {\bf q}},t)=\sum\limits_{j=1}^n A_{\nu,j}({\bf q,t}){\dot q}_j+b_\nu({\bf q},t)
\end{equation}
the condition (\ref{cetaev}) formally reproduces the standard way of treating linear kinematic constraints 
based on the definition of virtual displacements 
\begin{equation}
	\label{virtdispllin}
	\sum\limits_{j=1}^n A_{\nu,j}\delta q_j=0
\end{equation}
(indeed $\partial g_\nu/\partial {\dot q}_j= A_{\nu,j}$) which is present in the main treatises on kinematic constraints (it suffices to mention \cite{neimark}).
Obviously (\ref{virtdispllin}) is (\ref{virtdisplol}) when (\ref{vinclin}) is (\ref{ec}).

\noindent
A debated topic which is at the center of our discussion concerns precisely the fact of accepting the condition (\ref{cetaev}) even in the general case of kinematic constraints (\ref{gvinc}), even nonlinear. 
The postulate (\ref{cetaev}) is known in the literature as the ${\check {\rm C}}$etaev condition, 
this is why we placed the superscript $(c)$.

\begin{rem}
The condition (\ref{cetaev}) is not certainly the only way of establishing the virtual displacements in terms of the functions (\ref{gvinc}): a second way which has received attention in the literature requires the introduction of the variations $\delta {\dot q}_j$ and overall the virtual displacements must verify the request
		\begin{equation}
		\label{displvak}
		\delta^{(v)} g_\nu =\sum\limits_{j=1}^n \left(
		\dfrac{\partial g_\nu}{\partial q_j}\delta q_j+
		\dfrac{\partial g_\nu}{\partial {\dot q}_j}\delta {\dot q}_j
		\right)=0.
	\end{equation}
The superscript $(v)$ refers to ``vakonomic'' approach, which has the theoretical foundations in 
\cite{kozlov1}-- \cite{kozlov3}.
Although we will not deal with condition (\ref{displvak}), we point out some points of comparison between the two conditions. First of all, the holonomic case (\ref{vincol}) falls within (\ref{displvak}) only setting
	$g_\nu=f_\nu$ in (\ref{displvak}), so that (\ref{virtdisplol}) is obtained, since $\delta {\dot q}_j=0$. 
	On the other hand, for holonomic constraints the direct definition $g_\nu=f_\nu$ in (\ref{cetaev})
	(rather than the derivative (\ref{ec})) would not make sense. In other words, the two formulas match in the holonomic case $f=f({\bf q,t})$ as long as 
	$$
	\delta^{(v)}f_\nu=\delta^{(c)}{\dot f}_\nu.
	$$
	A second difference emerges in the case of linear kinematic constraints (\ref{vinclin}): on the one hand $\delta^{(c)}g_\nu=0$ is (\ref{virtdispllin}), on the other 
	one in the linear case (\ref{vinclin}) condition (\ref{displvak}) writes
	\begin{equation}
		\label{displvaklin}
		\delta^{(v)}g_\nu=
		\sum\limits_{i,j=1}^n \left( \dfrac{\partial A_{\nu,i}}{\partial q_j}{\dot q}_i
		+\dfrac{\partial b_\nu}{\partial q_j}\right)\delta q_j+\sum\limits_{j=1}^n A_{\nu,j}\delta {\dot q}_j=0
	\end{equation}
	which is apparently very different.
\end{rem}

\subsection{The equations of motion}

\noindent
We briefly report the standard method that is used to determine the corresponding equations of motion which correspond to hypothesys  (\ref{cetaev}).
The situation where the two equations (\ref{dapl1}) and (\ref{cetaev}) are coupled is straightforward: as (\ref{cetaev}) shows, the whole set of virtual displacements $\delta {\bf q}$ builds the vector space orthogonal to the space generated by $\frac{\partial g_\nu}{\partial {\dot {\bf q}}}$, $\nu=1, \dots, k$; since the latter vectors are independent, their linear combinations cover the totality of the vectors orthogonal to all the $\delta {\bf q}$. Among them we find 
the vector in round brackets in (\ref{dapl}) so that it is necessarily 
\begin{equation}
\label{eq1specie}
\dfrac{d}{dt}\dfrac{\partial {\cal L}}{\partial {\dot q}_j}-\dfrac{\partial {\cal L}}{\partial q_j}
-F^{(j)}_{NP}=\sum\limits_{\nu=1}^k \lambda_\nu \dfrac{\partial g_\nu}{\partial {\dot q}_j}, \qquad j=1, \dots, n
\end{equation}
with $\lambda_\nu$ unknown multiplying factors. The just written equations have to be coupled with the constraint equations (\ref{gvinc}) to forma a system of $n+k$ equations in the $n+k$ unknowns ${\bf q}$, $\lambda_\nu$, $\nu=1, \dots, k$. The constraint forces $(R^{(1)}, \dots, R^{(n)})$ are smoooth according to 
(\ref{iclagr}) and the case of (\ref{dapl1}) is analogous.

\noindent
The possibility of writing the equations of motion without the multipliers (uncomfortable unknown quantities)
passes through the explicit writing of the conditions (\ref{cetaev}) which in an abstract and general way we indicate by
\begin{equation}
\label{betanuj}
\sum\limits_{j=1}^n \beta_{\nu,j}\delta q_j=0, \qquad \nu=1, \dots, k
\end{equation}
\noindent
where $\beta_{\nu,j}$ are the elements of a $k\times n$ matrix with full rank $k$. 
Without losing generality we can determine the displacements as a function of the first independent displacements $\delta q_1$, $\dots$, $\delta q_m$, $m=n-k$:
\begin{equation}
	\label{lmexpl}
	\delta q_{m+\nu}=\sum\limits_{r=1}^m \gamma_{\nu,r}\delta q_r, \qquad \nu=1, \dots, k
\end{equation}
where $m=n-k$ and  $\delta q_1, \dots, \delta q_m$ are arbitrary. By replacing (\ref{lmexpl}) in (\ref{dapl1}) for each $\nu=1, \dots, k$ and taking into account that 
$\delta q_1$, $\dots$, $\delta q_m$ are independent parameters, the equations of motion can be written as

\begin{equation}
	\label{eq2specie}
		\dfrac{d}{dt}\dfrac{\partial {\cal L}}{\partial {\dot q}_r}-\dfrac{\partial {\cal L}}{\partial q_r}
		+\sum\limits_{\nu=1}^k \gamma_{\nu,r}\left( 
		\dfrac{d}{dt}\dfrac{\partial {\cal L}}{\partial {\dot q}_{m+\nu}}-\dfrac{\partial {\cal L}}{\partial q_{m+\nu}}\right)
		=F^{(r)}_{NP} + \sum\limits_{\nu=1}^k 
		\gamma_{\nu,r}F^{(m+\nu)}_{NP}\qquad r=1, \dots, m
\end{equation}
to be combined with (\ref{gvinc}), this time too. The $m+k=n$ equations contain the $n$ unknown functions $q_1, \dots, q_n$: in addition to the lower number of equations, the obvious advantage of the second system (\ref{eq1specie}) with respect to (\ref{eq1specie}) is the absence of the multipliers $\lambda_\nu$.

\noindent
When (\ref{betanuj}) assumes the form (\ref{cetaev}), that is $\beta_{\nu,j}=\frac{\partial g_\nu}{\partial {\dot q}_j}$, then the functions appearing in (\ref{lmexpl}) are 
\begin{equation}
	\label{lambdanur}
\gamma_{\nu, r}=\dfrac{\partial \alpha_\nu}{\partial {\dot q}_r}, \qquad \nu=1, \dots, k, \;\;r=1, \dots, m
\end{equation}
with
\begin{equation}
	\label{veldip}
\alpha_\nu=	{\dot q}_{m+\nu}(q_1, \dots, q_n, {\dot q}_1, \dots, {\dot q}_m,t), \qquad \nu=1, \dots, k
\end{equation}
deduced from (\ref{gvinc}) and by virtue of the condition of independence (except for renumerating the variables)
\begin{equation}
	\label{vincind}
	det\left(\dfrac{\partial g_\nu}{\partial {\dot q}_{m+s}}\right)\not =0 \qquad \nu,s=1, \dots, k.
\end{equation}
The relations (\ref{veldip} assign to ${\dot q}_1$, $\dots$, ${\dot q}_m$ the role of independent velocities and to the remaining ones ${\dot q}_{m+1}$, $\dots$, ${\dot q}_n$ that of dependent velocities.
In order to verify (\ref{lambdanur}) simply calculate the derivatives of $g_\nu=0$ with respect to the independent velocities ${\dot q}_r$, $r=1, \dots, m$:
$$
\dfrac{\partial g_\nu}{\partial {\dot q}_r}+\sum\limits_{s=1}^k \dfrac{\partial g_\nu}{\partial {\dot q}_{m+s}}
\dfrac{\partial \alpha_s}{\partial {\dot q}_r}=0\;\;\Rightarrow\;\;
\sum\limits_{r=1}^m\left(\dfrac{\partial g_\nu}{\partial {\dot q}_r}\delta q_r+\sum\limits_{s=1}^k \dfrac{\partial g_\nu}{\partial {\dot q}_{m+s}}
\dfrac{\partial \alpha_s}{\partial {\dot q}_r}\delta q_r\right)=0
$$
and the comparison with (\ref{cetaev}) leads to 
\begin{equation}
	\label{virtdispdip}
	\delta q_{m+\nu}=\sum\limits_{r=1}^m \dfrac{\partial \alpha_\nu}{\partial {\dot q}_r}\delta q_r, \qquad \nu=1, \dots, k
\end{equation}
where the virtual variations $\delta q_1$, $\dots$, $\delta q_r$ are independent. Consequently the equations of motion (\ref{eq2specie}) assume the form
\begin{equation}
	\label{vnl0}
	\dfrac{d}{dt}\dfrac{\partial {\cal L}}{\partial {\dot q}_r}-\dfrac{\partial {\cal L}}{\partial q_r}
	+\sum\limits_{\nu=1}^k\dfrac{\partial \alpha_\nu}{\partial {\dot q}_r}
	\left( \dfrac{d}{dt}\dfrac{\partial {\cal L}}{\partial {\dot q}_{m+\nu}}-
	\dfrac{\partial {\cal L}}{\partial q_{m+\nu}}\right)
	=F^{(r)}_{NP}+\sum\limits_{\nu=1}^k \dfrac{\partial \alpha_\nu}{\partial {\dot q}_r} F^{(m+\nu)}_{NP}, 
	\qquad r=1,\dots, m.
\end{equation}

\section{The ${\check {\rm C}}$etaev rule}

\subsection{Is the  ${\check {\rm C}}$etaev assumption inferable?}

\noindent
The rule (\ref{cetaev})	seems to have no real theoretical justification and the anomaly of matching the displacements to the kinetic derivatives is frequently pointed out. The same condition seems to be supported by the practical confirmation of the correct equations of motion and largely accepted.

\noindent	
Due to the axiomatic role of (\ref{cetaev}), our attention has fallen on an claimed demonstration \cite{flan} starting from the constraint equations and which we are going to comment on.

\noindent
The proof in question (Section III, B of \cite{flan}) can be formulated as follows.
Recovering the expression of ${\dot g}_\nu$ in the diagram just above taking into account the partition among independent and dependent variables as in (\ref{vincind}), we can write
$$
\sum\limits_{j=1}^n \dfrac{\partial g_\nu}{\partial q_j}{\dot q}_j+
\sum\limits_{i=1}^m \dfrac{\partial g_\nu}{\partial {\dot q}_i}{\q2dot^{..}}_i+
\sum\limits_{s=1}^ k\dfrac{\partial g_\nu}{\partial {\dot q}_{m+s}}{\q2dot^{..}}_{m+s}+
\dfrac{\partial g_\nu}{\partial t}=0, \qquad \nu=1, \dots, k
$$
whence we deduce, by derivating with respect to ${\q2dot^{..}}_r$, $r=1, \dots, m$:
\begin{equation}
\label{dersec}
\dfrac{\partial g_\nu}{\partial {\dot q}_r}+\sum\limits_{s=1}^k \dfrac{\partial g_\nu}{\partial {\dot q}_{m+s}}
\dfrac{\partial {\q2dot^{..}}_{m+s}}{\partial {\q2dot^{..}}_r}=0, \qquad r=1, \dots, m
\end{equation}
Now, we point we the point we criticize in \cite{flan} states, adapting the symbols to our notations:
``Although the coordinate function $q_{m+s}=q_{m+s}(q_1, \dots, q_m,t)$, $s=1, \dots, k$ is unknown for 
nonintegrable (\ref{gvinc}), the dependent displacements 
\begin{equation}
\label{deltaflan}
\delta q_{m+s}=
\sum\limits_{i=1}^m \frac{\partial q_{m+s}}{\partial q_i}\delta q_i=
\sum\limits_{i=1}^m \frac{\partial {\dot q}_{m+s}}{\partial {\dot q}_i}\delta q_i=
\sum\limits_{i=1}^m 
\frac{\partial {\q2dot^{..}}_{m+s}}{\partial {\q2dot^{..}}_i}\delta q_i
\end{equation}
can be obtained in terms of the independent $\delta q_r$, $r-1, \dots, m$''.
The dubious step is, in our opinion, to misintepret what is not known with what cannot be obtained: 
as a matter of fact, the (known or unknown) dependence is ascribable only to geometric constraints, but it cannot mathematically be inferred from a nonintegrable differential expression (\ref{gvinc}).

\noindent
Getting back to (\ref{deltaflan}), it is clear that the characteristic condition of the holonomic case
given by the first equality (which is (\ref{virtdisplol} with explicit functions)
will lead to the condition (\ref{cetaev}) which holds, as we have already noticed in (\ref{ec}), in the holonomic case. Actually:
$$
\sum\limits_{j=1}^n \dfrac{\partial g_\nu}{\partial {\dot q}_j}\delta q_j=
\sum\limits_{i=1}^m\dfrac{\partial g_\nu}{\partial {\dot q}_i}\delta q_i+
\sum\limits_{s=1}^k \dfrac{\partial g_\nu}{\partial {\dot q}_{m+s}}\delta q_{m+s}=
\sum\limits_{i=1}^m \left(
\dfrac{\partial g_\nu}{\partial {\dot q}_i} +
\sum\limits_{s=1}^k \dfrac{\partial g_\nu}{\partial {\dot q}_{m+s}} 
\frac{\partial {\q2dot^{..}}_{m+s}}{\partial {\q2dot^{..}}_i}\right)\delta q_i=0
$$
The first step simply splits the variables into independent and dependent, the second step makes use of (\ref{deltaflan}), last equality, in order to rewrite $\delta q_{m+s}$. Finally, the entire expression is null owing to (\ref{dersec}).

\noindent
We stress that the key point is 
actually $\delta q_{m+k}=\sum\limits_{i=1}^m \frac{\partial q_{m+k}}{\partial q_i}\delta q_i$ which is
characteristic of holonomic systems and 
in principle not necessarily compatible with (\ref{cetaev}). Using such an identity in a non-proper way is, in our opinion, the point at which the reasoning falls.
This property in fact fots for holonomic systems, but for the latter we already know that ${\check {\rm C}}$etaev rule holds as an ordinary condition on displacements.

\noindent
We even more remark that, if the procedure were rigorous, the second derivatives would not even be needed: the reasoning is much simpler if we replace (\ref{dersec}) with
$$
\sum\limits_{i=1}^m 
\left( 
\dfrac{\partial g_\nu}{\partial {\dot q}_i}+\sum\limits_{s=1}^k \dfrac{\partial g_\nu}{\partial {\dot q}_{m+s}}
\dfrac{\partial {\dot q}_{m+s}}{\partial {\dot q}_i}\right)=0 \; \Rightarrow 
$$
and we argue that
$$
\sum\limits_{j=1}^n \dfrac{\partial g_\nu}{\partial {\dot q}_j}\delta q_j=
\sum\limits_{i=1}^m \dfrac{\partial g_\nu}{\partial {\dot q}_i}\delta q_i+
\sum\limits_{s=1}^k \dfrac{\partial g_\nu}{\partial {\dot q}_{m+s}}\delta q_{m+s}=
\sum\limits_{i=1}^m \left( \dfrac{\partial g_\nu}{\partial {\dot q}_i}+
\sum\limits_{s=1}^k \dfrac{\partial g_\nu}{\partial {\dot q}_{m+s}}
\frac{\partial {\dot q}_{m+s}}{\partial {\dot q}_i}\right)\delta q_i=0
$$
this time using the second equality in (\ref{deltaflan}). 

\noindent
The same argument is performed in \cite{flan}, Section III, C for higher order constraints of the type:
$$
h_\nu({\bf q}, {\dot {\bf q}}, \bfq2dot^{..},t)=0
$$
in order to prove the relation (analogous to (\ref{cetaev}))
\begin{equation}
\label{hvinc}
\sum\limits_{j=1}^n \dfrac{\partial h_\nu}{\partial {\q2dot^{..}}_j}\delta q_j=0.
\end{equation}
The derivative of the constraint equation makes the third derivative ${\h2dot^{...}}_\nu$ appear and the statement 
$$
\delta q_{m+s}=
\sum\limits_{i=1}^m \frac{\partial q_{m+s}}{\partial q_i}\delta q_i=
\sum\limits_{i=1}^m 
\frac{\partial {\q2dot^{...}}_{m+s}}{\partial {\q2dot^{...}}_i}\delta q_i
$$
analogous to (\ref{deltaflan}) and the same conclusions are claimed, to affirm the validation of (\ref{hvinc}).
Nevertheless in our opinion ``unknown'' is confused with ``not existing'' one more time, at the moment in which the dependent functions $q_{m+s}(q_1, \dots, q_m, t)$ have been hypothesized, although unknown.

\begin{rem}
The claimed proof of (\ref{cetaev}) does not actually require (\ref{displvak}) to hold, 
as one seems to understand from \cite{flan}: as a matter of fact, the reasoning starts from 
the differential expression ${\dot g}_\nu$ (the one preceding (\ref{dersec})), regardless of the condition in terms of displacements (\ref{displvak}).
\end{rem}

\noindent
The theoretical validation for a general constraint $g_\nu({\bf q}, {\dot {\bf q}},t)=0$ of the ${\check {\rm C}}$etaev rule is further supported, again according to \cite{flan}, by the fact that the same equations of motion induced by the rule and written in (\ref{eq1specie}) are claimed to derive from the Gauss Principle: we will not go into details of the mathematical reasoning that justifies the derivation from the principle, but we limit ourselves to observing that the validity of the identity
$$
\frac{\partial {\bf r}}{\partial q_j}=
\frac{\partial {\dot {\bf r}}}{\partial {\dot q}_j}=
\frac{\partial \bfr2dot^{..}}{\partial {\q2dot^{..}}_j}
$$ is invoked, which in our opinion can only be used in the holonomic case. We are going to deal precisely with the just written identity in the next Paragraph.

\subsection{An argument in support of the rule}

\noindent
Although the hypothesis (\ref{cetaev}) does not appear to have rigorous theoretical support, it is also widely considered in the literature for the formulation of motion and for the study of models; it also represents a very handy tool regarding the extension of the conditions for higher order constraints.
Here we are going to add a heuristic consideration which seems to us to support the validity of the condition (\ref{cetaev}).

\noindent
The matter can be formulated as follows: coming back to the formal notations of (\ref{newton}), we consider the standard variation
\begin{equation}
\label{deltars}
\delta {\bf r}_s = \sum\limits_{j=1}^n \dfrac{\partial {\bf r}_s}{\partial q_j}\delta q_r
\end{equation}
holding for a holonomic system; for the same system the known property $\dfrac{\partial {\dot {\bf r}}_s}{\partial {\dot q}_j}=
\dfrac{\partial {\bf r}_s}{\partial q_j}$ makes the previous formula equivalent to 
\begin{equation}
\label{deltarsdot}
\delta {\bf r}_s = \sum\limits_{j=1}^n \dfrac{\partial {\dot {\bf r}}_s}{\partial {\dot q}_j}\delta q_r
\end{equation}
since the generlized velocities ${\dot q}_1$, $\dots$, ${\dot q}_n$ are independent. In the nonholonomic case the addition of kinematic constraints does not allow to deduce the right virtual displacements from (\ref{deltars}), thus it is natural to to ask what relationship exists between (\ref{deltars}) and (\ref{deltarsdot}) in the nonholonomic case.
As a matter of fact, the definition of virtual displacements as in (\ref{cetaev}) is the only one that allows us to extend the equivalence also in the case of nonholonomic constraints, provided that the independent velocities ${\dot q}_r$, $r-1, \dots, m$ are considered:
\begin{propr}
	Assume that (\ref{cetaev}) holds for a set of $\nu=1, \dots, k$ kinematic constraints acting on the system ${\bf r}_s({\bf q},t)$, $s=1, \dots, N$. Then $\delta {\bf r}_s$ defined in (\ref{deltars}) verifies  
	\begin{equation}
	\label{virtdisp}
	\delta {\bf r}_s = \sum\limits_{r=1}^m \dfrac{\partial {\dot {\bf r}}_s}{\partial {\dot q}_r}\delta q_r, \qquad s=1, \dots, N
\end{equation}
where ${\dot q}_1$, $\dots$, ${\dot q}_m$ is the set of independent velocities as in (\ref{veldip}). 
\end{propr} 

\noindent
{\bf Proof}: Since, owing to (\ref{veldip}),
\begin{equation}
\label{velr}
{\dot {\bf r}_s}=\sum\limits_{r=1}^m\dfrac{\partial {\bf r}_s}{\partial q_r}{\dot q}_r +
\sum\limits_{\nu=1}^k \dfrac{\partial {\bf r}_s}{\partial q_{m+\nu}}\alpha_\nu + 
\dfrac{\partial {\bf r}_s}{\partial t}, \qquad s=1, \dots, N
\end{equation}
we see that 
$$
\dfrac{\partial {\dot {\bf r}}_s}{\partial {\dot q}_r}=
\dfrac{\partial {\bf r}_s}{\partial q_r}+
\sum\limits_{\nu=1}^k \dfrac{\partial {\bf r}_s}{\partial q_{m+\nu}}
\dfrac{\partial \alpha_\nu}{\partial {\dot q}_r}, \qquad s=1, \dots, N, \quad r=1, \dots, m
$$
hence 
\begin{equation}
	\label{virtdispind}
	 \sum\limits_{r=1}^m \dfrac{\partial {\dot {\bf r}}_s}{\partial {\dot q}_r}\delta q_r
	=
\sum\limits_{r=1}^m \left( \dfrac{\partial {\bf r}_s}{\partial q_r} +
\sum\limits_{\nu=1}^k \dfrac{\partial \alpha_\nu}{\partial {\dot q}_r}
\dfrac{\partial {\bf r}_s}{\partial q_{m+\nu}} \right) \delta q_r=
\sum\limits_{r=1}^m \dfrac{\partial {\bf r}_s}{\partial q_r} \delta q_r+
\sum\limits_{\nu=1}^k \dfrac{\partial {\bf r}_s}{\partial q_{m+\nu}} \underbrace{
	\sum\limits_{r=1}^m  \dfrac{\partial \alpha_\nu}{\partial {\dot q}_r}\delta q_r
}_{=\delta q_{m+\nu}} 
\end{equation}
by virtue of (\ref{virtdispdip}) and (\ref{virtdisp}) is proven. $\quad \square$

\noindent
The essential hypothesis has been (\ref{cetaev}), which implies (\ref{virtdispdip}), otherwise the property is not verified.

\noindent
It is interesting at this point to make a comparison between the velocity 
${\widehat {\dot {\bf r}}_s}$ compatible with the instantaneous configuration of the constrained system (inferred from (\ref{velr}) by blocking the time) and $\delta {\bf r}_s$: let us 
fill in the following scheme which we will comment on immediately afterwards.
$$
\begin{array}{lll}
\delta {\bf r}_s=\sum\limits_{r=1}^n\dfrac{\partial {\bf r}_s}{\partial q_r}\delta q_r+
\sum\limits_{\nu=1}^k\dfrac{\partial {\bf r}_s}{\partial q_{m+\nu}}\delta q_{m+\nu},  &
\delta q_{m+\nu}=\sum\limits_{r=1}^m\dfrac{\partial \alpha_\nu}{\partial {\dot q}_r}\delta q_r
& 
\Rightarrow\;\; 
\delta {\bf r}_s=
\sum\limits_{r=1}^m
\left(\dfrac{\partial {\bf r}_s}{\partial q_r}+
\sum\limits_{\nu=1}^k \dfrac{\partial {\bf r}_s}{\partial q_{m+\nu}}\sum\limits_{r=1}^m  
\dfrac{\partial \alpha_\nu}{\partial {\dot q}_r}\right)\delta  q_r\\
\\	
	{\widehat {\dot {\bf r}}_s}=
	\sum\limits_{r=1}^m \dfrac{\partial {\bf r}_s}{\partial q_r}{\dot q}_r+
	\sum\limits_{\nu=1}^k\dfrac{\partial {\bf r}_s}{\partial q_{m+\nu}}\alpha_\nu, &
{\dot q}_{m+\nu}=
\sum\limits_{r=1}^m\dfrac{\partial \alpha_\nu}{\partial {\dot q}_r}{\dot q}_r & 
\Rightarrow\;\; 
{\widehat {\dot {\bf r}}_s}=
\sum\limits_{r=1}^m
\left(\dfrac{\partial {\bf r}_s}{\partial q_r}+
\sum\limits_{\nu=1}^k \dfrac{\partial {\bf r}_s}{\partial q_{m+\nu}}\sum\limits_{r=1}^m  
\dfrac{\partial \alpha_\nu}{\partial {\dot q}_r}\right){\dot q}_r
\end{array}
$$
The first column simply consists of (\ref{deltars}) and the definition of virtual velocity. The upper entry of the second column contains (\ref{virtdispdip}), while the lower one corresponds to the hypothesis
\begin{equation}
\label{baraa}
	{\overline \alpha}_\nu (q_1, \dots, q_n, {\dot q}_1, \dots, {\dot q}_m,t):=
	\sum\limits_{r=1}^m \dfrac{\partial \alpha_\nu}{\partial {\dot q}_r}{\dot q}_r=
	\alpha_\nu\quad \textrm{for any} \quad \nu=1,\dots, k 	
\end{equation}
Finally, the third column makes the preceding elements up (the upper one is (\ref{virtdispind})), showing a perfect formal adherence of the two expressions. We summarize the main result in the following
\begin{prop}
Assume that (\ref{cetaev}) holds for a set of $\nu=1, \dots, k$ kinematic constraints (\ref{vinc}) on the system ${\bf r}_s({\bf q},t)$, $s=1, \dots, N$. 
If the request (\ref{baraa}) is fulfilled, then $\delta {\bf r}_s$ takes the same directions as the virtual velocities ${\widehat {\dot {\bf r}}_s}={\dot {\bf r}}_s -\partial {\bf r}_s/\partial t$ (see (\ref{velr})), as the independent generalized velocities ${\dot q}_1$, $\dots$, ${\dot q}_m$, $m=n-k$ vary.
\end{prop}

\noindent
We believe it appropriate to point out that the hypothesis (\ref{baraa}) corresponds to a very specific structure of the functions, according to the following

\begin{prop}
The functions $\alpha_\nu$ are homogeneous functions of first degree. This fact occurs if and only if the functions $g_\nu$ of (\ref{gvinc}) are all homogeneous of arbitrary degree and not necessarily of the same degree.
\end{prop}

\noindent
The first part of the statment is evident; the proof of the rest is in \cite{talhomo}.

\begin{exe}
Any holonomic constraint (\ref{vincol}) not depending on time $f_\nu({\bf q})=0$ verifies (\ref{baraa}) if written in the derivative form $g_\nu={\dot f}_\nu =\sum\limits_{j=1}^n \frac{\partial f_\nu}{\partial q_j}{\dot q}_j=0$ which is homogeneous of degree $1$. As an instance, for two points $P_1$ and $P_2$ at a constant distance $\ell$ the geometric constraint $(q_1-q_4)^2+(q_2-q_5)^2+(q_3-q_6)^2-\ell^2=0$ (where $(q_1,q_2,q_3)$ and $(q_4,q_5,q_6)$ are the cartesian coordinates of $P_1$ and $P_2$, respectively)
which is converted by 
the derivation to the homogeneous function of degree $1$ 
$(q_1-q_4)^2+({\dot q}_1-{\dot q}_4)+
(q_2-q_5)^2+({\dot q}_2-{\dot q}_5)+
(q_3-q_6)^2+({\dot q}_3-{\dot q}_6)=0$ 

\noindent
For a mobile constraint $f_\nu({\bf q},t)=0$ the same property is not true.
\end{exe}

\begin{exe}
The folloging constraints for two points $P_1$ and $P_2$
$$
\begin{array}{cll}
(i) &{\dot P}_1 \wedge {\dot P}_2={\bf 0} & \textrm{parallel velocities} \\  	
(ii) &{\dot P}_1 \cdot {\dot P_2}=0 & \textrm{perpendicular velocities}   \\
(iii)& |{\dot P}_1|= |{\dot P}_2| & \textrm{same magnitude of the velocities}  \\
(iv) & {\dot B}\wedge \overrightarrow{P_1 P_2}=0  &
\textrm{the velocity of the midpoint $B$ is parallel to the joining line}  \\
(v) & {\dot B}\cdot \overrightarrow{P_1 P_2}=0 & \textrm{the velocity of the midpoint $B$ is perpendicular to the joining line}  \\
(vi) & {\dot P}_1 \cdot\overrightarrow{P_1 P_2}={\dot P}_2 \cdot 
\overrightarrow{P_1 P_2}=0 &
\textrm{the velocities are perpendicular to the joining line} 
\end{array}
$$
fall into the category (\ref{baraa}): actually, fixing the lagrangian parameters as the cartesian coordinates $P_1=(q_1, q_2, q_3)$, $P_2=(q_4, q_5, q_6)$, the constraints just listed are written respectively

$$
\begin{array}{cl}
	(i) &{\dot q}_2{\dot q}_6-{\dot q}_3{\dot q}_5=0,\;
	{\dot q}_3{\dot q}_4-{\dot q}_1{\dot q}_6=0,\;
	{\dot q}_1{\dot q}_5-{\dot q}_2{\dot q}_4=0
	\\  	
	(ii) & {\dot q}_1{\dot q}_4+{\dot q}_2{\dot q}_5+{\dot q}_3{\dot q}_6=0 \\
	(iii)& \sqrt{{\dot q}_1^2+{\dot q}_2^2+{\dot q}_3^2}-
\sqrt{{\dot q}_1^2+{\dot q}_2^2+{\dot q}_3^2}=0
\\
	(iv) & \left\{
	\begin{array}{l}
	(q_3-q_6) ({\dot q}_2+{\dot q}_5)- (q_2-q_5) ({\dot q}_3+{\dot q}_6)=0\\
	(q_1-q_4) ({\dot q}_3+{\dot q}_6)- (q_3-q_6) ({\dot q}_1+{\dot q}_4)=0\\
	(q_2-q_5) ({\dot q}_1+{\dot q}_4)- (q_1-q_4) ({\dot q}_2+{\dot q}_5)=0
	\end{array}
\right.
		 \\
	(v) &(q_1-q_4) ({\dot q}_1+{\dot q}_4)+(q_2-q_5) ({\dot q}_2+{\dot q}_5)
	+(q_3-q_6) ({\dot q}_3+{\dot q}_6)=0\\
	(vi) & (q_1-q_2){\dot q}_1+(q_2-q_5){\dot q_2}+(q_3-q_6){\dot q}_3=0, \;
	(q_1-q_2){\dot q}_4+(q_2-q_5){\dot q}_5+(q_3-q_6){\dot q}_6=0
\end{array}
$$
They are all homogeneous functions of degree $1$ or $2$, hence the hypothesys (\ref{baraa}) is satisfied by the explicit functions $\alpha_\nu$, according to what we reported in Proposition $1$.
A more detailed analysis of the models is carried out in \cite{talmecc}.
\end{exe}

\noindent
The diagram just before (\ref{baraa}) highlights the most suitable context for the feasibility of (\ref{cetaev}): in the class of functions (\ref{baraa}) the arbitrariness of $\delta q_r$ and ${\dot q}_r$, $r=1, \dots, m$, means that the virtual displacements and the virtual velocities are the same vectors. Such a physical connotation, to say, is supported by at least two non-marginal issues:
\begin{itemize}
	\item[$(1)$] 
	The ideal constraint $\sum\limits_{s=1}^N {\bf R}_s \cdot \delta {\bf r}_s=\sum\limits_{j=1}^n {\cal R}^{(j)}\delta q_j =0$ (see (\ref{ic}) and (\ref{iclagr})) does actually represent the absence of virtual work $\sum\limits_{s=1}^N {\bf R}_s\cdot {\widehat {\dot {\bf r}}_s}=\sum\limits_{j=1}^n {\cal R}^{(j)}{\dot q_j} =0$ whenever (\ref{baraa}) holds: more precisely the following relation which can be easily deduced from the comparison of (\ref{eq1specie}) and (\ref{vnl0}) confirms the statement:
\begin{equation}
	\label{virtwork}
	\sum\limits_{j=1}^n{\cal R}^{(j)} {\dot q}_j = 
	\sum\limits_{\nu=1}^k
	(\alpha_\nu-{\overline \alpha}_\nu)
	\left(\dfrac{d}{dt}\dfrac{\partial {\cal L}}{\partial {\dot q}_{m+\nu}}-\dfrac{\partial {\cal L}}{\partial q_{m+\nu}}-F_{NP}^{(m+\nu)}\right).
\end{equation}
We see that the virtual work vanishes if whenever (\ref{baraa}) holds.
	\item[$(2)$] Consider the Hamiltonian function
	\begin{equation}
		\label{e}
		{\cal E}(q_1, \dots, q_n, {\dot q}_1, \dots, {\dot q}_n,t)=
		\sum\limits_{j=1}^n {\dot q}_j \dfrac{\partial {\cal L}}{\partial {\dot q}_j}-{\cal L}
	\end{equation}
and the one formulated by the only independent kinetic variables:
	\begin{equation}
	\label{estar}
	{\cal E}^*(q_1, \dots, q_n, {\dot q}_1, \dots, {\dot q}_m,t)= \sum\limits_{r=1}^m {\dot q}_r 
	\dfrac{\partial {\cal L}^*}{\partial {\dot q}_r}-{\cal L}^*.
\end{equation} 
where 
$$
		{\cal L}^*(q_1,\dots, q_n, {\dot q}_1, \dots, {\dot q}_m, t)
		={\cal L}(q_1, \dots, q_n, {\dot q}_1, \dots, {\dot q}_m, \alpha_1(\cdot), \dots, 
		\alpha_k(\cdot), t)
$$
is the Lagrangian function restricted to the independent velocities (see \ref{veldip}))
and $(\cdot)$ indicates $(q_1,\dots, q_n$, ${\dot q}_1, \dots, {\dot q}_m, t)$.
The two functions (\ref{e}) and (\ref{estar}) do not coincide when they are expressed in the same set of variables, i.~e.~$\left.{\cal E}\right\vert_{{\dot q}_{m+\nu}=\alpha_\nu}\not = {\cal E}^*$: the exact relation, which can be easily checked, is 
\begin{equation}
	\label{eestar}
		{\cal E}^*={\cal E}+
		\sum\limits_{\nu=1}^k({\overline \alpha}_\nu-\alpha_\nu)\dfrac{\partial {\cal L}}{\partial {\dot q}_{m+\nu}}.
\end{equation}		
Once again, we see that the hypothesis (\ref{baraa}) plays a unifying role in the definition of the physical quantities and (\ref{e}), (\ref{estar}) do coincide; this allows the energy of the system to be uniquely defined.
\end{itemize}

\section{Conclusion and next investigation}

\noindent
The class of displacements (\ref{cetaev}) sets linear conditions on the $\delta q_j$ which allow the d'A--L P.~to be used in order to establish the equations of motion even in the case of non-linear nonholonomic constraints.
The long--standing problem is whether it exists the possibility of deriving the set (\ref{cetaev}) from the constraint equations (\ref{gvinc}).
In our opinion the problem has not been resolved in \cite{flan} and the ${\check {\rm C}}$etaev rule remains
at least for the moment without a theoretical explanation.

\noindent
However, this circumstance does not detract from a series of positive aspects that the rule offers: the use of the d'A--L P.~makes the equations of motion written in a simple and direct way, the generalization to higher orders (see \ref{hvinc})) occurs through the $\delta q_j$ alone, therefore nothing changes from a formal point of view; many other advantageous aspects could be mentioned.

\noindent
A further advantage of the condition (\ref{cetaev}) combined with the d'A--L P.~is that it is not required to 
express about the possible commutation $\delta {\dot q}_j-\frac{d}{dt}{\delta q}_j$ (this is actually a debated question), which instead is indispensable whenever (\ref{displvak}) is assumed, necessarily combined with the integral formulation of a principle, owing to the presence of $\delta {\dot q}_j$. A stimulating and interesting analysis of the condition is performed in \cite{flan}.

\noindent
We also highlighted how the validity of (\ref{baraa}) (which, as reported in Proposition $1$, is characteristic of homogeneous constraints of any order) places the mechanical system in a decidedly natural physical context, where the displacements coincide with the virtual velocities, the virtual work of the constraint forces is zero and the energy of the system can be univocally defined (see (\ref{virtwork}), (\ref{eestar})).

\noindent
The conditions listed in Example $2$ show various nonholonomic constraints (some of them nonlinear) of systems that are certainly not marginal in the context of kinematic restrictions: 
This encourages us to think that (\ref{baraa}) actually covers a vast set of physically feasible nonholonomic constraints. From a mathematical point of view, the topic to be investigated is the possibility of formulating a zero level set through homogeneous functions.
Obviously the topic concerns stationary constraints and time-dependent ones requires a separate study.

\noindent
A second issue  we are investigating concerns the correlation between the two conditions (\ref{cetaev}) and (\ref{displvak}): if on the one hand it is simple to write the mathematical identity that links them (the so called transpositional relation), on the other it is not obvious to understand, for example, which category of systems admits both displacements, or what is the role of the commutation rule.
In recent decades the debate turns out to be animated in literature \cite{lemos}, \cite{li}, \cite{yang}.

\end{document}